\documentclass[manuscript,screen,sigconf]{acmart}
\AtBeginDocument{%
  \providecommand\BibTeX{{%
    \normalfont B\kern-0.5em{\scshape i\kern-0.25em b}\kern-0.8em\TeX}}}

\setcopyright{acmcopyright}
\copyrightyear{2023}
\acmYear{2023}
\acmDOI{XXXXXXX.XXXXXXX}

\acmConference[WYSD '23]{Workshop YOUR study design!  Participatory critique and refinement of participants' studies '23, ACM/IEEE International Conference on Human-Robot Interaction '23 (HRI'23), March 13--16, 2023}{March 13--16, 2023}{Stockholm, SE}
\begin{document}

\title{Evaluating Gesture Recognition in Virtual Reality}


\author{Sandeep Reddy Sabbella}
\email{sabbella@diag.uniroma1.it}
\orcid{0000-0003-2629-1734}
\affiliation{%
  \institution{Sapienza Università di Roma}
  \city{Rome}
  \country{Italy}
}

\author{Sara Kaszuba}
\email{kaszuba@diag.uniroma1.it}
\orcid{0000-0001-6201-1512}
\affiliation{%
  \institution{Sapienza Università di Roma}
  \city{Rome}
  \country{Italy}
}

\author{Francesco Leotta}
\email{leotta@diag.uniroma1.it}
\orcid{0000-0001-9216-8502}
\affiliation{%
  \institution{Sapienza Università di Roma}
  \city{Rome}
  \country{Italy}
}

\author{Pascal Serrarens}
\email{pse@pale.blue}
\affiliation{%
  \institution{PaleBlue}
  \city{Stavanger}
  \country{Norway}
}

\author{Daniele Nardi}
\email{nardi@diag.uniroma1.it}
\orcid{0000-0001-6606-200X}
\affiliation{%
  \institution{Sapienza Università di Roma}
  \city{Rome}
  \country{Italy}
}

\renewcommand{\shortauthors}{Sabbella et al.}

\begin{abstract}
Human-Robot Interaction (HRI) has become increasingly important as robots are being integrated into various aspects of daily life. One key aspect of HRI is gesture recognition, which allows robots to interpret and respond to human gestures in real-time. Gesture recognition plays an important role in non-verbal communication in HRI. To this aim, there is ongoing research on how such non-verbal communication can strengthen verbal communication and improve the overall efficiency of the system, thereby enhancing the user experience with the robot. However, there are several challenges that need to be addressed in gesture recognition systems which include data generation, transferability, scalability, generalizability, standardization, and lack of benchmarking of the gestural systems. In this preliminary paper, we want to address the challenges of data generation using virtual reality simulations and the issues of standardization by presenting gestures to some commands that can be used as a standard in ground robots.
\end{abstract}

\ccsdesc[500]{Human-centered computing~HCI design and evaluation methods}
\ccsdesc[500]{Human-centered computing~Virtual reality}
\ccsdesc[500]{Human-centered computing~Gestural input}
\ccsdesc[500]{Computing methodologies~Interactive simulation}
\ccsdesc[100]{Computing methodologies~Simulation environments}

\keywords{Human-Robot Interaction, Virtual Reality, Gesture Recognition, Virtual Datasets, Virtual Human Avatars}



\maketitle

\section{Introduction}

Human-Robot Interaction (HRI) is a research field of interest in recent years. In general, HRI aims to study how people interact with robots, how to design robots interacting with people naturally and intuitively, and to create robots that can seamlessly integrate into human environments and work alongside people to achieve common goals. While verbal Communication accounts for 80\% of general interactions, non-verbal communication, such as gestures and facial expressions can convey a lot of information about a person's intentions, emotions, and attitudes. Gesture recognition is crucial in HRI, allowing robots to understand and respond to non-verbal human cues. It enables robots to comprehend people's gestures, e.g., pointing, and facial expressions, and use this information to guide their actions in various applications. In agriculture, a human can require the robot to do specific tasks when busy or speaking with fellow workers.

Out of several challenges in gesture recognition, data is one of the most important challenges to be addressed. Even though several datasets are available in the domain of sign language, actual feasible data for gesture generation in HRI is scarce. One way to generate such gestural datasets is using Virtual Reality (VR) simulators. In a VR simulation, a character can perform multiple gestures with several variations to the virtual environment. This allows generating a large dataset along with the ground truth, which can be used to train and evaluate gesture recognition approaches. Our research interest in exploring this novel approach is to discover if the gestures generated in a VR simulation can match the performance of the gestures performed in the real world and to answer "how efficiently can the model trained on such data perform in real-world data?".

\section{Related Works}
The recent advent of deep learning-based techniques such as Convolutional Neural Networks (CNNs) and Recurrent Neural Networks (RNNs) have been shown to be effective at recognizing gestures in real-time. Some of such approaches that have been proposed for sign recognition include 2D CNN Models \cite{Tasmere,Yu}, 3D CNN Models\cite{s19245429,SINGH202176,8743159}, GANs\cite{Tang,Suzuki}, Spatio-temporal networks and RNN Models\cite{Samaan,Lai}. These systems have been reported to have high accuracy in gesture recognition tasks. Still, the accuracy may change depending on the data's amount and quality, and the model's specific implementation. Recently, transformers and their efficiency in speech and vision domains caught the attention of researchers. 

In recent years, researchers have made significant progress in developing new techniques to recognize, generate, and animate hand and body gestures in VR. With the advances in virtual avatars, several reviews \cite{Jiang,Zhang} were conducted on Human-Avatar Interaction (HAI) in virtual environments, highlighting the importance of gesture-based communication and rehabilitation. They discussed various approaches for animating avatars, including motion capture, keyframe animation, and physics-based animation. Generation of human-like avatars from images was presented in article\cite{LI202181}, which involves learning a generative model of the 3D avatar from a large dataset of motion-captured videos. Ultimately, the authors demonstrated their method's effectiveness in generating realistic and diverse avatars for different poses and actions.

\section{Approaches and Challenges of Gesture recognition}

The state-of-the-art methodologies in gesture recognition techniques for HRI include various techniques broadly classified into Computer Vision-based, Machine Learning-based, Wearable-based, Hybrid, and Virtual Reality-based techniques. Computer Vision-based techniques process images or videos of a person performing a gesture and extract features that can be used to recognize the gesture. Machine Learning-based techniques train models that can recognize gestures from image or video data. Wearable-based methods use sensors such as accelerometers or gyroscopes to detect and track gestures. Hybrid techniques combine multiple input sensor modalities to improve the accuracy and robustness of gesture recognition systems. Virtual Reality-based techniques use VR simulations to generate datasets of gestures that can be used to train and evaluate gesture recognition systems.

However, several challenges need to be addressed in gesture recognition systems for HRI \cite{Carfì}. These challenges include variations in gesture performance, occlusions and self-occlusions, background noise and lighting, real-time processing, privacy-security concerns, scalability and generalizability, and benchmarking and standardization. Ensuring that gesture recognition systems can accurately detect and interpret human gestures in different scenarios and environments while protecting the privacy and ensuring compliance with regulations remains a significant challenge for researchers and developers in the field of HRI.

Addressing these challenges requires interdisciplinary collaborations among the advances in computer vision, machine learning, and robotics and a better understanding of the human factors involved in HRI. Hopefully, most of these can be solved with the use of simulated virtual human avatars and will be detailed in the upcoming section.

\section{Gesture Definitions and Data Generation}
There were several definitions of gestures based on the applications and their usage context, yet most still need to be standardized. Regarding industrial applications, gesture definitions change according to the type of industry and context used. General movement and context-specific gestures were defined to generalize and standardize the signs in the industry. These were defined after following some gestures for communicative purposes \cite{Jia_chuan} in HRI and from the gesture language between human-human and human-robot and in the workspace for the CANOPIES project\footnote{https://canopies.inf.uniroma3.it/}. The primary scientific goals of CANOPIES involve creating innovative approaches to HRI, the collaboration between humans and robots (HRC), and coordination among multiple robots (MRC). These methodologies aim to establish a productive collaborative system between human workers and teams of multiple robots to achieve precision farming of permanent crops. The effectiveness of these approaches will be evaluated in the setting of table-grape vineyards. This is the reason for collecting the data in the table-grapes field to understand the noise and the variations in lighting conditions under the grape canopy. 

\begin{figure}[]
  \includegraphics[width=0.98\columnwidth, height= 11.4cm]{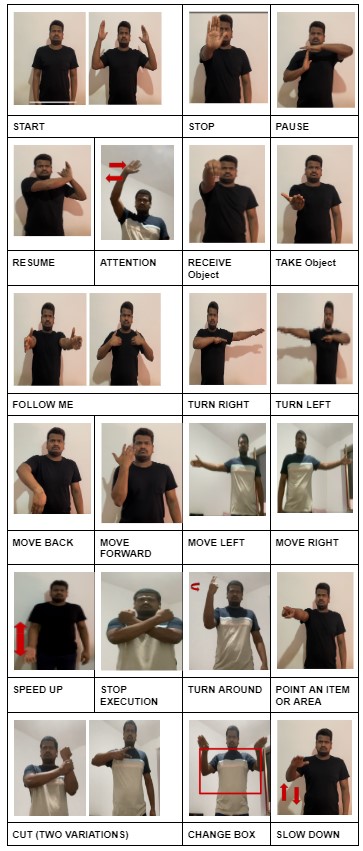}
  \caption{Gestures defined and their descriptions.}
  \Description{Defined gestures based on the workspace ethic and human communication.}
  \label{gesturedef}
\end{figure}

The defined gestures, as shown in Figure \ref{gesturedef} consist of 10 dynamic and 11 static signs\footnote{We initially defined large set of gestures for acquisition purposes. They can be conveniently reduced to a smaller set based on performance and feedback from experimentation}, where 9 gestures can be performed using two hands while the rest are with one hand. Some gesture definitions were specific to the Project of CANOPIES [\textit{CUT, CHANGE BOX, RECEIVE, TAKE}], while the rest are generic to the all ground robots. Human workers in the table-grape field will be working along with robots in performing agricultural activities, which require at least one hand in most cases to instruct the robot, without interrupting their activities. The 9 gestures that need two hands can have any other tools in the hands of the human in most scenarios.

\subsection{Real-time Data}

\begin{figure}[]
  \includegraphics[width=0.9\columnwidth, height=3.8cm]{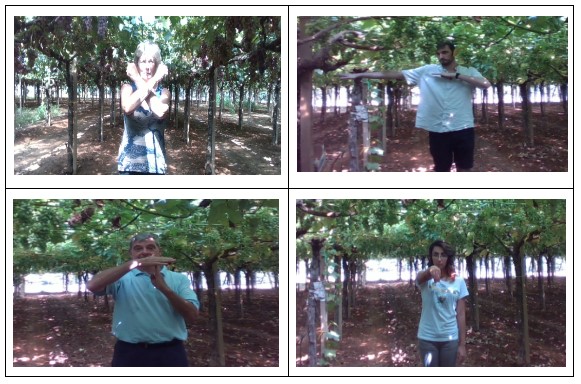}
  \caption{Real-time Gestures captured in the table-grape field with different persons at different distances.}
  \Description{Real-time Gestures captured in the table-grape field with different persons at different distances.}
  \label{realdata}
\end{figure}

The real-time data was captured using Intel realsense Depth camera D435i with IMU and stereo vision\cite{intel-realsense} in the table-grapes field in Aprilia (Lazio), a worksite for the project. The gesture data was recorded after obtaining consent to use and publish it for research purposes. To this aim, 8 persons from different age groups and gender participated in recording the defined gestures, as shown in Figure \ref{realdata}. These gestures were recorded at the resolution 1280x720 at 30 FPS with depth RGB-D information. The gestures were performed and captured at 4 different distance settings, starting at 1.5 meters from the camera and increasing the interval by considering multiples of 1.5, to store such information in RGB-D Ros-bags.

All 21 gestures were recorded into a single ROS-Bag file for each distance per person. All the files were then post-processed to extract videos of the performed signs based on the type of the gesture. Then, Mediapipe's pose estimation \cite{BlazePose} was applied to verify if all the keypoints of the poses were being detected. 

\subsection{VR-Simulation for Gesture Data}

One way to generate datasets of gestures is through VR simulations. Here, a character (Human-Avatar) can perform gestural movements in a virtual environment, and the system can record the gestures and their corresponding annotations. This allows for the generation of a large number of examples of different gestural signs, which can be used to train and evaluate gesture recognition systems. Indeed, a single character can be mapped to many animations and the virtual environment can be configured accordingly to the right parameters.

There are a few advantages of using VR to generate a dataset:
\begin{itemize}

\item VR can provide a controlled environment where the lighting, background noise, and other factors can be standardized to minimize the issues that would be faced in real-world datasets, such as occlusions and background noise.

\item VR simulations can provide a safe and repeatable way to generate a large number of examples of different gestures, without the need for expensive or dangerous equipment.

\item VR simulations can provide a way to generate examples of gestures that would be difficult or impossible to perform in the real world, such as dangerous or demanding ones.
\end{itemize}

\subsection{Virtual Simulated Data}
Autodesk character creator was used to generate 10 characters representing the varied height, gender, ethnicity, and age. These characters were imported to Unity. The collected videos were processed using DeepMotion to extract animations and then were mapped to the generated characters in the Unity. DeepMotion uses a neural network to extract the AI-based animation, motion tracking, and physics simulation to generate a full-body avatar that tracks motion from video, a single camera, or a minimal set of VR sensors. These tracked motions are then used to reconstruct the animation flow, and the extracted key points of the joints serve as a base reference to the joint's movement from the captured video. 


We have verified if the animations were working as expected by manually playing characters with respective motions in the simulation. There was a human-in-loop to evaluate the animations with respect to the actual videos by playing them simultaneously. There was no effective metric for this evaluation so the effectiveness of the animation is measured through human observation. Still, any animations that don't perform well after the extraction from the respective video were discarded and re-recorded from different angles, which can capture the movement optimally to generate the proper animation. By doing so, we have estimated that we can generate extensive data for gesture recognition. To this aim, we believe in successfully deploying a deep learning algorithm trained on the hybrid data of virtually simulated and real-time data acquired.

\begin{figure}[]
  \includegraphics[width=0.85\columnwidth, height = 6.8cm]{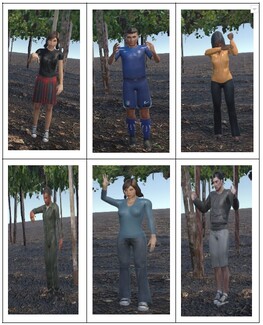}
  \caption{Virtual Human Avatars performing animated gestures in the table-grape field simulation.}
  \Description{Virtual Human Avatars performing animated gestures in the table-grape field simulation.}
  \label{VHAdata}
\end{figure}

\section{Experimentation}
The experimentation will be conducted in three Phases to answer the research question, "How do the models trained on Hybrid data perform in real-world?". The first phase involves 20 people participating in performing the defined gestures. Each user will perform all the gestures with 4 repetitions of different distances from the camera. All these gestures will be recorded using the Intel realsense D435i \cite{intel-realsense} or a similar camera to be consistent with the previously registered data. The recorded data will be post-processed to videos/frames to feed into the training network.

The second experimentation phase generates more virtually simulated data by creating additional virtual human characters. These characters will be mapped to the extracted gesture animations. The system will record the gestures performed by these characters through multiple cameras in the Unity tool with various distances and angles. We repeat the data collection with one character performing gestures and multiple characters performing the same or different gestures. Once the data generation is completed, we will employ 70-30\% for Virtual - to - Real data to generate a hybrid dataset.

The Third phase of the experimentation uses Mediapipe's pose estimation \cite{BlazePose} to evaluate the generated data. The recorded frames will be evaluated to check if the key points are being recognized and at what threshold. This will be taken as a base score. In the case of virtual simulated data, the identified key-points will be evaluated using Unity-generated joint coordinates as the ground truth. Our experiments and evaluations are based on the hypothesis that a model trained on the hybrid data can be used to identify the gestures both in simulation and the real-world. A suitable deep-learning approach will be adopted to train a network on such data. The network will be trained only on Real-data, Virtual-simulated data, and hybrid data. The Mediapipe's model and the trained models will be compared against each other with variations of training data. After evaluations, we aim to find the possible data combinations (virtual-real) for improving the network's performance. Hence, the network with the best performances will be deployed to a virtual robot in the simulation, being able to see the virtual human avatar performing gestures so that the virtual robot will communicate or execute the corresponding actions based on the gesture commands given by the virtual character. The users in the simulation will have control of choosing gestures via a keyboard (input device) in a \textit{non-immersive} experience and with the hand controllers in an \textit{immersive} experience based on the device. All these experiments will further be cataloged for system enhancements, and the model's performance evaluation will be conducted on real-world data by deploying such an approach into the real robot.

\section{Questions and Discussion Points}
The goal of this study design is to get answers for the following questions.
\begin{enumerate}
\item[Q1.] What are the most effective proportions of simulated vs. real data for training a model?
\item[Q2.] Can we use a single model to evaluate the gestures in simulation and real-world?
\item[Q3.] How efficiently can the model trained on such data perform in real-world data?
\item[Q4.] What are evaluation metrics that can be employed for gesture recognition in the context of HRI, and how can they be addressed?

\end{enumerate}

 \section{Conclusions}
VR simulation aims to solve the difficulty of acquiring real data in the wild. There are many advantages to using virtual simulated data with a degree of control over noise, lighting, and scale. Cost and time consumption for the initial system set-up limits many users from using VR simulations to generate gesture datasets. Along with cost, a person should have expertise in using such tools since learning them could be expensive. To this aim, there is active research for generating characters with deep learning networks such as GAN, but it is still in its infancy.
Overall, while generating datasets of gestures through VR simulations can be a helpful technique for training and evaluating gesture recognition systems, it is essential to keep in mind the limitations of this approach and validate the performance of the gesture recognition system in real-world scenarios. We plan on comparing the performances of trained networks with data obtained in the lab vs. VR-generated data.


\bibliographystyle{ACM-Reference-Format}
\bibliography{bibliography}


\end{document}